\documentclass[journal=jacsat,manuscript=article]{achemso}

\usepackage[version=3]{mhchem} 
\usepackage[T1]{fontenc}
\usepackage{siunitx}
\usepackage{graphicx}
\usepackage[utf8]{inputenc}
\DeclareUnicodeCharacter{2212}{-}
\usepackage{underscore}
\usepackage{hyperref}

\author{Sanchali Mitra}
\email{sanchali_mitra@sutd.edu.sg}
\affiliation{Science, Mathematics and Technology, Singapore University of Technology and Design (SUTD), 8 Somapah Road, Singapore 487372, Singapore}
\author{Santanu Mahapatra}
\affiliation{Nano-Scale Device Research Laboratory, Department of Electronic Systems Engineering,
Indian Institute of Science (IISc) Bangalore, Bangalore 560012, India
}
\email{santanu@iisc.ac.in}

\title[An \textsf{achemso} demo]
  {A follow-up on the sulphur atom popping model for MoS$_2$ memristor}

\abbreviations{IR,NMR,UV}

\begin{document}


\begin{abstract}
  The mechanism of resistive switching in two-dimensional (2D) semiconductor-based memristors is intriguing, and our conventional knowledge of bulk-oxide based memristors does not apply to these devices. Experimental data indicate that the genesis of resistive switching may be intrinsic to the 2D semiconducting active layer, as well as resulting from the movement of electrode atoms. Employing reactive-force field (ReaxFF) molecular dynamics simulations, we introduced the ``sulphur atom popping model'' [npj 2D Mater. Appl. 5, 33 (2021)] to elucidate the intrinsic nature of non-volatile resistive switching in 2D molybdenum disulfide-based memristors. In this paper we provide additional perspective to this model using density functional theory. We also discuss the limitations of universal machine learning interatomic potentials in reproducing ReaxFF simulation results.
\end{abstract}

\section{Main}

Two-dimensional (2D) semiconductors have emerged as promising candidates for the active layer in metal-insulator-metal based non-volatile resistive memory, also known as memristors \cite{li_resistive_2024}. The origin of this notable resistive switching has been debated over; whether this phenomenon is inherent to the 2D material or if the electrode materials contribute. We introduced a hypothesis termed ``S atom popping'' \cite{mitra_theory_2021} to elucidate this behavior in monolayer MoS$_2$-based memristors. We adopted reactive-force-field based (ReaxFF) \cite{senftle_reaxff_2016} based molecular dynamics (MD) simulations and ab initio calculations to formulate the theory. A recent article by Shah et al.\cite{shah_revisiting_2024} questions the validity of this model. Although the authors reproduced some results from the original article \cite{mitra_theory_2021} using Reaxff-based molecular dynamics simulations, they also show that these results are not in agreement when simulations are performed with M3GNet, \cite{chen_universal_2022} a universal machine learning interatomic potential (MLIP). They also claimed that the popped-state cannot be stabilized using ab-initio calculations. In this paper, we first provide a commentary on the many technical flaws of this work. We then present additional density-functional-theory (DFT) based calculations to reinforce the ``S atom popping'' model.

\subsection{Charge equilibration in molecular dynamics simulations}

The primary goal of MD simulation in these works is to monitor the movement of atoms in response to an applied electric field with the notion that under an electric field \textbf{E}, an atom with a charge of \textbf{q} experiences a force of \textbf{q}$\cdot$\textbf{E} and thus its trajectory is influenced accordingly. Although MLIPs are getting popular nowadays, M3GNet has neither been trained for nor can predict atomic charges \cite{chen_universal_2022}. The only quantities it can predict from a structure object (such as \href{https://pymatgen.org/pymatgen.core.html}{Pymatgen Structure} or  \href{https://wiki.fysik.dtu.dk/ase/ase/atoms.html}{ASE atoms})  are energy, forces, and stress. Incorporating atomic charge information in MLIP is a grand challenge and only few like: CHGNet \cite{deng_chgnet_2023} and SpookyNet \cite{unke_spookynet_2021} (developed mostly for molecular systems) offer such capacity, albeit CHGNet predicts magnetic moments of the atoms, not the charge or electronegativity directly; the magnetic moments however, can be converted to charge using  \href{https://github.com/CederGroupHub/chgnet/blob/main/chgnet/utils/vasp_utils.py}{a heuristic-based algorithm}. It is important to emphasize that simply assigning pre-calculated atomic charges to the system that was statically derived at the beginning, perhaps by a first principles-based method, is not enough for the present problem, the charges should be dynamically updated within the MLIP at every MD step as done in ReaxFF for a meaningful simulation. In the ``Methods/Electric field coupled molecular dynamics'' section the authors write ``The partial charges (q) on all atoms were calculated dynamically at each step of MD using the charge equilibration (QEq) scheme of Rappe and Goddard'', yet they seem unaware of the fact that M3GNet is inherently unable to compute or assign atomic charges dynamically. The QEq charge equilibration scheme\cite{rappe_charge_1991,nakano_parallel_1997,rick_dynamical_1994} uses an electronegativity equalization principle, which requires recalculating per-atom charges at every timestep to remain consistent with evolving bonding environments. In a reactive system, whenever a bond breaks or a new bond forms the charges of the participating atoms usually change significantly and thus should be updated immediately for meaningful dynamics under an electric field. If an MLIP does not support charge assignment, in theory, QEq cannot be applied directly. We are not sure how the simulations progress in this case, most probably with fixed charges assigned at the beginning of the simulations. Ideally, the QEq method should raise an error if no updated charges are found in the force field. In any case, the results obtained would be erroneous for the intended usage. Thus M3GNet simulations \cite{shah_revisiting_2024} conducted under finite electric field to track the trajectories of the atoms, do not provide any \textit{scientifically valid} findings.

\subsection{ReaxFF versus M3GNET}

The authors \cite{shah_revisiting_2024} question the reliability of the Mo-S ReaxFF force field introduced by Ostadhossein et al. \cite{ostadhossein_reaxff_2017} using quite strong wordings. However, in truth, this force field is exclusively designed for 2D or layered MoS$_2$ with defects, and the introductory paper demonstrates rigorous validations with structures containing various types of defects that agree well with experimental findings. Subsequently, this force field has been used many times to elucidate experimental findings for intricate MoS$_2$ systems \cite{carmesin_quantum-dot-like_2019, patra_defect_2018, hong_computational_2017}  that affirm its accuracy.

On the contrary, it is well known that the accuracy of ML potentials is strongly dependent on the quality and quantity of the training data\cite{unke_machine_2021}.  The M3GNet universal ML interatomic potential was trained on 62783 different materials and 188349 frames/structures in total, obtained from the Materials project\cite{jain_commentary_2013}. However, the \href{https://figshare.com/articles/dataset/MPF_2021_2_8/19470599}{training dataset} has only 14 materials (39 frames in total) that contain only the species Mo and S [mpids according to the dataset are: mp-1627, mp-673645, mp-990083, mp-2815, mp-1018809, mp-1434, mp-1023939, mp-1025874, mp-2164, mp-1023924, mvc-7052, mp-31257, mvc-11780, mp-558544]. Some of these entries have been deprecated, and most have been updated in the current version of the Materials Project ( \href{https://legacy.materialsproject.org}{Legacy Materials project} versus \href{https://next-gen.materialsproject.org}{Next-gen Materials Project}). A visual inspection reveals that all these materials are bulk materials except one, and none of them contain any S vacancy, the study of which was the main objective of this work. In addition, from the  \href{https://matbench-discovery.materialsproject.org}{Matbench Discovery leaderboard} \cite{riebesell_matbench_2023} we don’t find M3GNet (as on September 19, 2025), within the top 10.

The system under study is quite intricate, a vacancy containing 3-layer 2D material sandwiched between two electrodes. The notion that universal MLIPs trained on pristine bulk materials can accurately predict the properties of such a system is highly ambitious. The MLIPs trained on bulk data must be used with utmost caution when applied to intricate systems such as surfaces, interfaces, 2D materials, and defective materials. The recently released MLIP MatterSim\cite{yang_mattersim_2024}, built based on the M3GNet architecture itself and trained on millions (1/5/17, different models) of materials data, issues a warning about this on their \href{https://github.com/microsoft/mattersim}{GitHub repo}; ``MatterSim-v1 is designed specifically for atomistic simulations of bulk materials. Applications or interpretations beyond this scope should be approached with caution. For instance, when using the model for simulations involving surfaces, interfaces, or properties influenced by long-range interactions, the results may be qualitatively accurate but are not suitable for quantitative analysis. In such cases, we recommend fine-tuning the model to better align with the specific application.''  Therefore, the strong claim made by the authors that the universal ML force field M3GNet (trained on 14 Mo-S materials, no vacancy structures) is more ``rigorous'' and can provide more accurate results than the ReaxFF potential that was specifically designed for this system and purpose and has since been validated many times, is unrealistic in our opinion. The comparison between M3GNET and ReaxFF is irrelevant, since M3GNET is not a charge-aware force field. ReaxFF is specifically designed for reactive systems, where frequent bond formation and breaking lead to substantial charge redistribution.

\subsection{Ab-initio Simulations}

\subsubsection{Geometry Optimization and structural stability}

In the section ``Supercell Optimization \& Total Energy Analysis'' the authors \cite{shah_revisiting_2024} stated: ``It was observed that all the relaxed supercells from ReaxFF had the S atom in the popped state (Fig. 1b) (Supplementary Fig. 1). However, for M3GNet and DFT, the relaxed supercells had the S atom returned to the parent state (Fig. 1a).'' It should be noted that the popped state, as reported in the original article \cite{mitra_theory_2021}, was obtained by applying an electric field for an extended period during MD simulations. This process not only alters the atomic positions but also induces charge redistribution within the system. Additionally, MD simulations cause shifts in the in-plane positions of neighbouring atoms.

\begin{figure}
    \centering
    \includegraphics[scale=0.75]{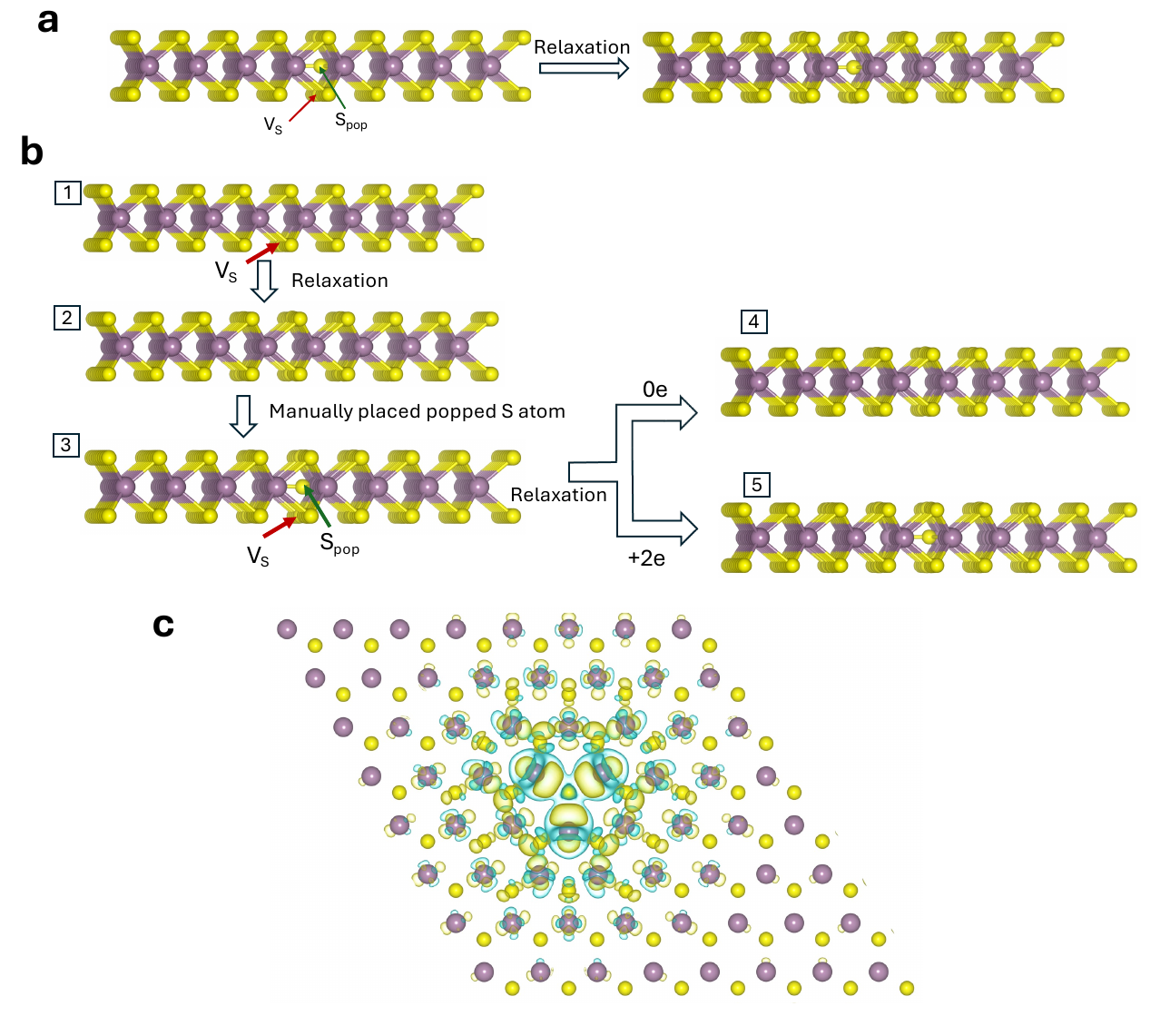}
    \caption{\textbf{a} Relaxation of an 8$\times$8 supercell of monolayer MoS$_2$, where a popped state is created by manually moving the S atom opposite to the vacancy to the Mo plane. After relaxation, the S atom remains within the Mo layer. \textbf{b} An 8$\times$8 supercell of monolayer MoS$_2$ containing an S vacancy (1) the vacancy structure is first relaxed (2), then a popped state is created by manually positioning an S atom in the Mo plane (3), the popped state is then relaxed without adding extra electrons, (4) and relaxed again with two extra electrons added, (5) without extra electrons, the popped state returns to the original S plane, whereas with extra electrons, the popped atom remains in the Mo plane. \textbf{c} Isosurface plot of the charge density difference between popped state stabilized MoS$_2$ layer with two added electrons and popped state stabilized MoS$_2$ layer without added electrons. The isosurface level is set as 0.0007 eV\r{A}$^{-3}$.}
    \label{relax}
\end{figure}

We have conducted following experiments to reproduce their observations. We constructed an 8×8 supercell of a monolayer MoS$_2$ with a single sulphur vacancy and then explored the geometry relaxation of the popped state using two different approaches. In the first approach, we manually moved the S atom opposite to the vacancy to the Mo-plane and performed a relaxation step (Figure 1a). In the second approach, we first relaxed the vacancy-containing structure, then manually positioned the S atom in the Mo-plane of the relaxed structure, followed by another relaxation step (Figure 1b). The density functional theory (DFT) calculations were carried out using the (Projected Augmented Waves) PAW \cite{blochl_projector_1994} basis set and the GGA-PBE (Generalized Gradient Approximation of Perdew-Burke-Ernzerhof) \cite{perdew_generalized_1996} exchange-correlation functional as implemented in the Vienna Ab initio Simulation Package (VASP) \cite{kresse_efficiency_1996} \cite{kresse_ultrasoft_1999} with gamma-point sampling. Our results show that in the first case, the S atom remains in the Mo plane after relaxation, whereas in the second case, it reverts to its original position This difference arises because, in the second approach, the initial relaxation of the parent vacancy state modifies the positions of neighboring atoms compared to the first approach, altering the forces acting on the manually placed S atom. Thus, the authors’ experimentation with the popped state relaxation is inconclusive. In this context, we further assert that filament structure obtained by ReaxFF simulation in bulk-oxide based memristors \cite{onofrio_atomic_2015}, might collapse under DFT relaxation, and thus does not disprove the formation of filaments \cite{yang_observation_2012}.

\subsubsection{Nudge elastic band calculations}

In the section ``Supercell Optimization \& Total Energy Analysis,'' the authors\cite{shah_revisiting_2024} state that they were unable to stabilize the popped state using DFT relaxation. However, in the following section, ``Minimum Energy Path (MEP) from Parent to Popped State,'' they presented Climbing Image Nudged Elastic Band (CINEB) calculations between the vacancy state and the popped state.
NEB calculations are usually performed between two stable states to discover the transition states, and the physical significance of performing NEB calculations between a stable and an “unstable” state is highly debatable. Additionally, in Section ``Supercell Optimization \& Total Energy Analysis,'', the authors stated that ``MoS$_2$ supercells with popped states have a higher energy by about $\sim$ 8.1 eV than parent states for all supercell sizes.'' However, during their NEB calculations, the popped state energy was reported to be only $\sim$ 2.6 eV higher than the vacancy state (Figure 3a \cite{shah_revisiting_2024}). The authors did not provide any explanation for this substantial $\sim$ 5.5 eV energy drop in the popped state during NEB calculations. The stark inconsistency in their own statements casts doubt on the dependability of their DFT-NEB simulations.

\subsubsection{Importance of charge localization under electric-field}

It is well known that S vacancies in MoS$_2$ introduce mid-gap states capable of capturing and trapping electrons \cite{lee_electrical_2023}\cite{li_sulfur_2017}. When a popped state is created, it effectively introduces another S vacancy, which can further enhance electron trapping, especially in the presence of an electric field. To further validate the popped atom mechanism, we applied a static electric field to Structure 1 (Figure 1a), where the popped atom remains stable. We also conducted CINEB calculations \cite{henkelman_climbing_2000} (Figure 2) in a similar way. Our results show that, in the absence of an electric field, the popped state has an energy approximately 2.74 eV higher than the vacancy state (Figure 2a), with NEB calculations placing the popped state at the highest energy level (Figure 2b). However, when an electric field is applied, the energy of the popped state changes significantly, while the energy of the vacancy state remains largely unchanged (Figure 2a). As the field strength increases, the energy of the popped state eventually drops below that of the vacancy state.  Additionally, under an electric field, an energy valley appears during the CINEB simulations (Figure 2b). The intermediate states between the vacancy and the popped state show no significant energy change, highlighting that the effect is specific to the popped state.  If the popped state were merely an artifact of ReaxFF, field-coupled DFT simulations would not have exhibited such a pronounced energy difference with the ground-state DFT. The charge density difference plot (Figure 2c) reveals that the application of an electric field induces charge localization around the popped state.  The associated decrease in total energy under the field suggests that the charge localization around the popped atom makes it energetically more stable.

\begin{figure}
    \centering
    \includegraphics[scale=0.6]{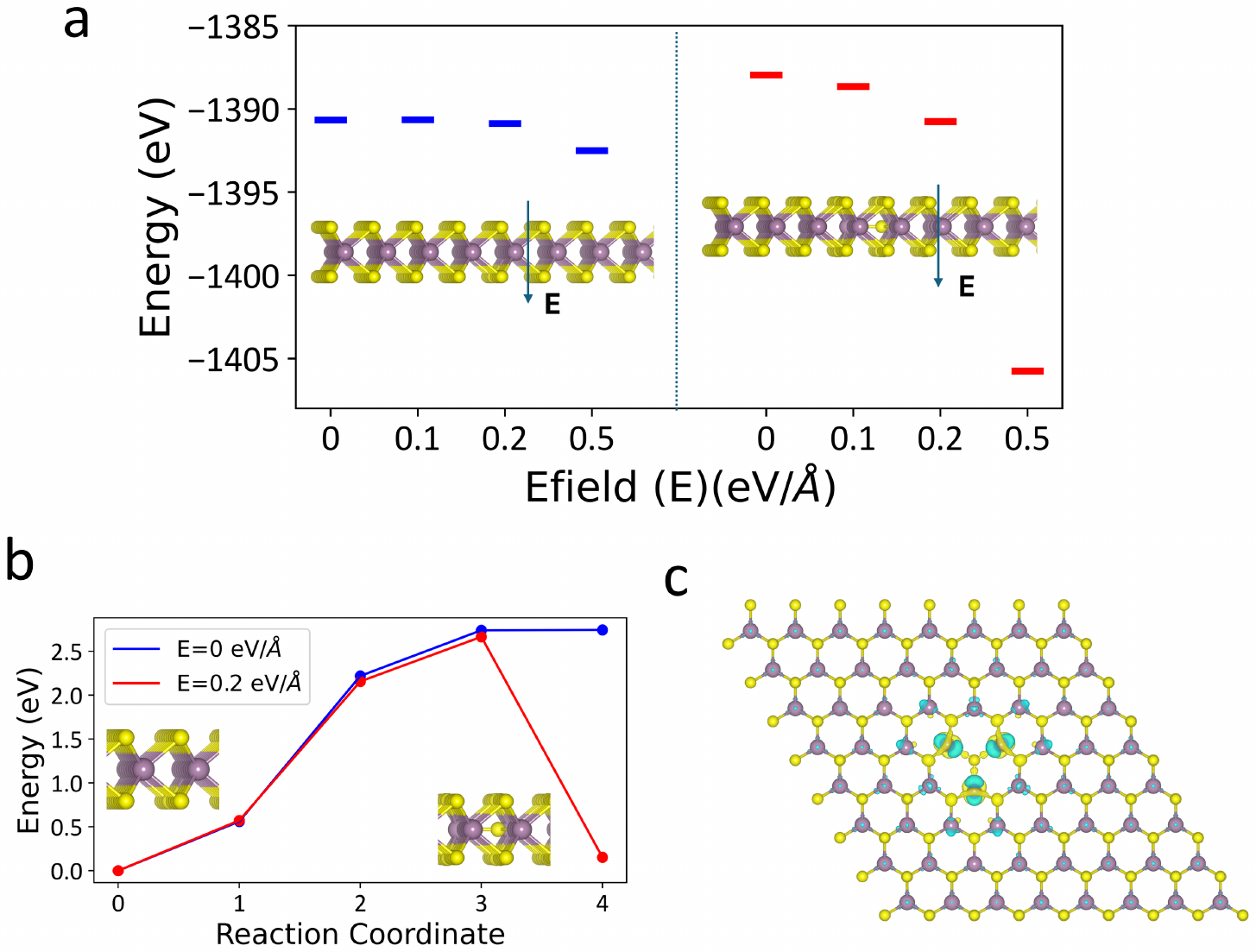}
    \caption{\textbf{a} Plot of the total energy values for the vacancy and popped structures under a static electric field (E) applied along the negative z-direction. The left panel (blue) represents the energy values of the parent vacancy states, while the right panel (red) shows the energy values of the popped states as the electric field increases. \textbf{b} CINEB calculation for S atom popping without an electric field (blue) and with an electric field (red). \textbf{c} Isosurface plot of the charge density difference between E=0.5 eV/\r{A} and E=0 eV/\r{A}.  The isosurface level is set as 0.0004 eV\r{A}$^{-3}$.}
    \label{neb}
\end{figure}

\subsubsection{Non-volatility of popped-state}

However, in DFT, transferring the charge distribution from an electric field-induced simulation to a zero-field simulation is highly challenging. Hence, once the electric field is removed, the self-consistent field (SCF) procedure in DFT reverts the system to the charge-neutral ground state, effectively removing the trapped charges near the popped site. As depicted in Fig 1f of Ref. \cite{mitra_theory_2021}, S atom popping causes the appearance of new states at and around the Fermi level, along with two additional localized defect states within the bandgap. Under an applied electric field, electrons can occupy these defect states, become trapped, and consequently alter the charge state of the defects. Ground-state DFT simulations, which maintain charge neutrality by fixing the total number of electrons, cannot inherently capture this electron-trapping phenomenon. Without explicitly adding extra electrons, all electrons are either paired or delocalized, leaving no excess electrons to occupy the defect-induced trap states. As a result, these states remain empty, and the defect retains its neutral charge state. Therefore, it is essential to introduce additional electrons into the system to accurately model electron trapping in ground-state DFT\cite{komsa_native_2015}. This approach accelerates the electron redistribution process, mimicking the effects that would naturally occur under an external electric field.  

We repeat our earlier relaxation process of the popped state (Figure 1b), now by introducing two extra electrons into the system. Our results showed that after relaxation, the popped sulfur atom remained within the Mo plane (Figure 1b). Further analysis of the charge density in the presence of extra electrons revealed that the added electrons were not uniformly distributed across the MoS$_2$ layer and instead became localized around the popped atom (Figure 1c). It was found that the popped state was unstable with one additional electron but attains stability with the addition of two or more electrons. This is because increasing the number of electrons alters charge localization around the popped state; and with only one extra electron, the localization is insufficient to stabilize it. 
To investigate this charge-induced stability further, we conducted ab initio molecular dynamics (AIMD) simulations as implemented in VASP. The simulations were first performed within the NPT ensemble at 300 K with two additional electrons. We observe that popped sate becomes unstable after 0.5ps. However, this timescale is much higher than reported by Shah et al.\cite{shah_revisiting_2024} and gives a clear indication than trapped charged around the popped state strengthen its stability. We redo the simulation with adding one more electron (i.e. total three) to the system. This time we find the popped state to be stable for 1ps. We then raised the temperature at 500 K and found the popped state to be stable for 3ps.  This observation reinforce the role of charge localization in stabilizing it though we cannot predict the amount of electron to be added \textit{a priori}. However, under real-time conditions, prolonged exposure to an electric field naturally redistributes electrons around the popped state, further enhancing its stability. For the same reason, charge neutral NEB calculations, conducted by Shah et al.\cite{shah_revisiting_2024} cannot be regarded as a definitive method for validating the existence of a charge-dependent state.

One may question why ReaxFF does not require extra electrons to stabilize the popped state, while DFT does. The reason lies in the fundamental differences between the two methods. DFT is a quantum-mechanical framework that determines the ground-state or lowest-energy configuration of a system from its electron density. In contrast, ReaxFF is a classical reactive force field parameterized against DFT data, designed to capture bond formation and breaking through its bond-order formalism. The popped state is not a true ground-state structure, but rather an excited configuration induced by an electric field. Ground-state DFT alone cannot reproduce such field-induced charge redistribution unless explicit external constraints, such as additional electrons, are applied. In fact, many studies in defect physics routinely account for charge states of defects by modifying the number of valence electrons in DFT simulations \cite{komsa_native_2015,noh_stability_2014,freysoldt_first-principles_2014, prasad_charge_2023}. ReaxFF, however, does not strictly enforce the ground state description. Instead, it can model the dynamic changes and charge redistribution that occur in an excited or non-equilibrium state, such as the popped state under an electric field. As a result, ReaxFF relaxation requires no extra electrons and can directly capture the charge localization around the popped atom.

\section{Summary}
Modeling such an electric-field-induced pop state at the DFT level is inherently complex. It is crucial to reiterate that the ``popped state'' is derived from a dynamic simulation, after prolonged exposure to a certain amount of electric field. In contrast, DFT relaxation aims to identify a system's ground-state configuration under the specified conditions. Simply placing an atom in a ``popped'' position fails to account for the crucial charge redistribution that occurs during the dynamic popping process, which is essential for its stability. Therefore, attempting to check the stability of a manually created popped state in a ground-state DFT simulation does not disprove its existence.  It instead highlights the limitations of ground-state methods in capturing field-induced configurations.  

In summary, it is infeasible to replicate all observations from ReaxFF-based molecular dynamics simulations within the confines of static DFT calculations. In fact, reactive force fields are built to handle large-scale, time-dependent reactive simulations that are not computationally feasible with DFT.  However, in this present study, we reinforce our ``sulphur atom popping model'' with two complimentary demonstrations: (i) electric field-coupled NEB, energetics, and corresponding charge redistribution, and (ii) stabilisation of the popped state via charge trapping using static relaxation and AIMD simulations. Together, these results strengthen our initial findings from the ReaxFF simulations and provide a consistent physical picture.

\begin{acknowledgement}

The authors are thankful to Dr. Arnab Kabiraj of the Agency for Science, Technology, and Research (A$*$STAR), Singapore, for insightful input on the machine learning interatomic potential.

\end{acknowledgement}

\bibliography{2DMem}

\providecommand{\latin}[1]{#1}
\makeatletter
\providecommand{\doi}
  {\begingroup\let\do\@makeother\dospecials
  \catcode`\{=1 \catcode`\}=2 \doi@aux}
\providecommand{\doi@aux}[1]{\endgroup\texttt{#1}}
\makeatother
\providecommand*\mcitethebibliography{\thebibliography}
\csname @ifundefined\endcsname{endmcitethebibliography}
  {\let\endmcitethebibliography\endthebibliography}{}
\begin{mcitethebibliography}{32}
\providecommand*\natexlab[1]{#1}
\providecommand*\mciteSetBstSublistMode[1]{}
\providecommand*\mciteSetBstMaxWidthForm[2]{}
\providecommand*\mciteBstWouldAddEndPuncttrue
  {\def\EndOfBibitem{\unskip.}}
\providecommand*\mciteBstWouldAddEndPunctfalse
  {\let\EndOfBibitem\relax}
\providecommand*\mciteSetBstMidEndSepPunct[3]{}
\providecommand*\mciteSetBstSublistLabelBeginEnd[3]{}
\providecommand*\EndOfBibitem{}
\mciteSetBstSublistMode{f}
\mciteSetBstMaxWidthForm{subitem}{(\alph{mcitesubitemcount})}
\mciteSetBstSublistLabelBeginEnd
  {\mcitemaxwidthsubitemform\space}
  {\relax}
  {\relax}

\bibitem[Li \latin{et~al.}(2024)Li, Chen, Wang, Niu, Xu, Miao, and
  Li]{li_resistive_2024}
Li,~X.; Chen,~N.; Wang,~B.; Niu,~M.; Xu,~M.; Miao,~X.; Li,~X. Resistive
  {Memory} {Devices} at the {Thinnest} {Limit}: {Progress} and {Challenges}.
  \emph{Advanced Materials} \textbf{2024}, \emph{36}, 2307951\relax
\mciteBstWouldAddEndPuncttrue
\mciteSetBstMidEndSepPunct{\mcitedefaultmidpunct}
{\mcitedefaultendpunct}{\mcitedefaultseppunct}\relax
\EndOfBibitem
\bibitem[Mitra \latin{et~al.}(2021)Mitra, Kabiraj, and
  Mahapatra]{mitra_theory_2021}
Mitra,~S.; Kabiraj,~A.; Mahapatra,~S. Theory of nonvolatile resistive switching
  in monolayer molybdenum disulfide with passive electrodes. \emph{npj 2D
  Materials and Applications} \textbf{2021}, \emph{5}, 33\relax
\mciteBstWouldAddEndPuncttrue
\mciteSetBstMidEndSepPunct{\mcitedefaultmidpunct}
{\mcitedefaultendpunct}{\mcitedefaultseppunct}\relax
\EndOfBibitem
\bibitem[Senftle \latin{et~al.}(2016)Senftle, Hong, Islam, Kylasa, Zheng, Shin,
  Junkermeier, Engel-Herbert, Janik, Aktulga, Verstraelen, Grama, and
  Van~Duin]{senftle_reaxff_2016}
Senftle,~T.~P.; Hong,~S.; Islam,~M.~M.; Kylasa,~S.~B.; Zheng,~Y.; Shin,~Y.~K.;
  Junkermeier,~C.; Engel-Herbert,~R.; Janik,~M.~J.; Aktulga,~H.~M.;
  Verstraelen,~T.; Grama,~A.; Van~Duin,~A. C.~T. The {ReaxFF} reactive
  force-field: development, applications and future directions. \emph{npj
  Computational Materials} \textbf{2016}, \emph{2}, 15011\relax
\mciteBstWouldAddEndPuncttrue
\mciteSetBstMidEndSepPunct{\mcitedefaultmidpunct}
{\mcitedefaultendpunct}{\mcitedefaultseppunct}\relax
\EndOfBibitem
\bibitem[Shah \latin{et~al.}(2024)Shah, Dar, and
  Shrivastava]{shah_revisiting_2024}
Shah,~A.~A.; Dar,~A.~B.; Shrivastava,~M. Revisiting the origin of non-volatile
  resistive switching in {MoS2} atomristor. \emph{npj 2D Materials and
  Applications} \textbf{2024}, \emph{8}, 80\relax
\mciteBstWouldAddEndPuncttrue
\mciteSetBstMidEndSepPunct{\mcitedefaultmidpunct}
{\mcitedefaultendpunct}{\mcitedefaultseppunct}\relax
\EndOfBibitem
\bibitem[Chen and Ong(2022)Chen, and Ong]{chen_universal_2022}
Chen,~C.; Ong,~S.~P. A universal graph deep learning interatomic potential for
  the periodic table. \emph{Nature Computational Science} \textbf{2022},
  \emph{2}, 718--728\relax
\mciteBstWouldAddEndPuncttrue
\mciteSetBstMidEndSepPunct{\mcitedefaultmidpunct}
{\mcitedefaultendpunct}{\mcitedefaultseppunct}\relax
\EndOfBibitem
\bibitem[Deng \latin{et~al.}(2023)Deng, Zhong, Jun, Riebesell, Han, Bartel, and
  Ceder]{deng_chgnet_2023}
Deng,~B.; Zhong,~P.; Jun,~K.; Riebesell,~J.; Han,~K.; Bartel,~C.~J.; Ceder,~G.
  {CHGNet} as a pretrained universal neural network potential for
  charge-informed atomistic modelling. \emph{Nature Machine Intelligence}
  \textbf{2023}, \emph{5}, 1031--1041\relax
\mciteBstWouldAddEndPuncttrue
\mciteSetBstMidEndSepPunct{\mcitedefaultmidpunct}
{\mcitedefaultendpunct}{\mcitedefaultseppunct}\relax
\EndOfBibitem
\bibitem[Unke \latin{et~al.}(2021)Unke, Chmiela, Gastegger, Schütt, Sauceda,
  and Müller]{unke_spookynet_2021}
Unke,~O.~T.; Chmiela,~S.; Gastegger,~M.; Schütt,~K.~T.; Sauceda,~H.~E.;
  Müller,~K.-R. {SpookyNet}: {Learning} force fields with electronic degrees
  of freedom and nonlocal effects. \emph{Nature Communications} \textbf{2021},
  \emph{12}, 7273\relax
\mciteBstWouldAddEndPuncttrue
\mciteSetBstMidEndSepPunct{\mcitedefaultmidpunct}
{\mcitedefaultendpunct}{\mcitedefaultseppunct}\relax
\EndOfBibitem
\bibitem[Rappe and Goddard(1991)Rappe, and Goddard]{rappe_charge_1991}
Rappe,~A.~K.; Goddard,~W.~A. Charge equilibration for molecular dynamics
  simulations. \emph{The Journal of Physical Chemistry} \textbf{1991},
  \emph{95}, 3358--3363\relax
\mciteBstWouldAddEndPuncttrue
\mciteSetBstMidEndSepPunct{\mcitedefaultmidpunct}
{\mcitedefaultendpunct}{\mcitedefaultseppunct}\relax
\EndOfBibitem
\bibitem[Nakano(1997)]{nakano_parallel_1997}
Nakano,~A. Parallel multilevel preconditioned conjugate-gradient approach to
  variable-charge molecular dynamics. \emph{Computer Physics Communications}
  \textbf{1997}, \emph{104}, 59--69\relax
\mciteBstWouldAddEndPuncttrue
\mciteSetBstMidEndSepPunct{\mcitedefaultmidpunct}
{\mcitedefaultendpunct}{\mcitedefaultseppunct}\relax
\EndOfBibitem
\bibitem[Rick \latin{et~al.}(1994)Rick, Stuart, and Berne]{rick_dynamical_1994}
Rick,~S.~W.; Stuart,~S.~J.; Berne,~B.~J. Dynamical fluctuating charge force
  fields: {Application} to liquid water. \emph{The Journal of Chemical Physics}
  \textbf{1994}, \emph{101}, 6141--6156\relax
\mciteBstWouldAddEndPuncttrue
\mciteSetBstMidEndSepPunct{\mcitedefaultmidpunct}
{\mcitedefaultendpunct}{\mcitedefaultseppunct}\relax
\EndOfBibitem
\bibitem[Ostadhossein \latin{et~al.}(2017)Ostadhossein, Rahnamoun, Wang, Zhao,
  Zhang, Crespi, and Van~Duin]{ostadhossein_reaxff_2017}
Ostadhossein,~A.; Rahnamoun,~A.; Wang,~Y.; Zhao,~P.; Zhang,~S.; Crespi,~V.~H.;
  Van~Duin,~A. C.~T. {ReaxFF} {Reactive} {Force}-{Field} {Study} of
  {Molybdenum} {Disulfide} ({MoS}$_{\textrm{2}}$ ). \emph{The Journal of
  Physical Chemistry Letters} \textbf{2017}, \emph{8}, 631--640\relax
\mciteBstWouldAddEndPuncttrue
\mciteSetBstMidEndSepPunct{\mcitedefaultmidpunct}
{\mcitedefaultendpunct}{\mcitedefaultseppunct}\relax
\EndOfBibitem
\bibitem[Carmesin \latin{et~al.}(2019)Carmesin, Lorke, Florian, Erben, Schulz,
  Wehling, and Jahnke]{carmesin_quantum-dot-like_2019}
Carmesin,~C.; Lorke,~M.; Florian,~M.; Erben,~D.; Schulz,~A.; Wehling,~T.~O.;
  Jahnke,~F. Quantum-{Dot}-{Like} {States} in {Molybdenum} {Disulfide}
  {Nanostructures} {Due} to the {Interplay} of {Local} {Surface} {Wrinkling},
  {Strain}, and {Dielectric} {Confinement}. \emph{Nano Letters} \textbf{2019},
  \emph{19}, 3182--3186\relax
\mciteBstWouldAddEndPuncttrue
\mciteSetBstMidEndSepPunct{\mcitedefaultmidpunct}
{\mcitedefaultendpunct}{\mcitedefaultseppunct}\relax
\EndOfBibitem
\bibitem[Patra \latin{et~al.}(2018)Patra, Zhang, Schulman, Chan, Cherukara,
  Terrones, Das, Narayanan, and Sankaranarayanan]{patra_defect_2018}
Patra,~T.~K.; Zhang,~F.; Schulman,~D.~S.; Chan,~H.; Cherukara,~M.~J.;
  Terrones,~M.; Das,~S.; Narayanan,~B.; Sankaranarayanan,~S. K. R.~S. Defect
  {Dynamics} in 2-{D} {MoS}$_{\textrm{2}}$ {Probed} by {Using} {Machine}
  {Learning}, {Atomistic} {Simulations}, and {High}-{Resolution} {Microscopy}.
  \emph{ACS Nano} \textbf{2018}, \emph{12}, 8006--8016\relax
\mciteBstWouldAddEndPuncttrue
\mciteSetBstMidEndSepPunct{\mcitedefaultmidpunct}
{\mcitedefaultendpunct}{\mcitedefaultseppunct}\relax
\EndOfBibitem
\bibitem[Hong \latin{et~al.}(2017)Hong, Krishnamoorthy, Rajak, Tiwari, Misawa,
  Shimojo, Kalia, Nakano, and Vashishta]{hong_computational_2017}
Hong,~S.; Krishnamoorthy,~A.; Rajak,~P.; Tiwari,~S.; Misawa,~M.; Shimojo,~F.;
  Kalia,~R.~K.; Nakano,~A.; Vashishta,~P. Computational {Synthesis} of
  {MoS}$_{\textrm{2}}$ {Layers} by {Reactive} {Molecular} {Dynamics}
  {Simulations}: {Initial} {Sulfidation} of {MoO}$_{\textrm{3}}$ {Surfaces}.
  \emph{Nano Letters} \textbf{2017}, \emph{17}, 4866--4872\relax
\mciteBstWouldAddEndPuncttrue
\mciteSetBstMidEndSepPunct{\mcitedefaultmidpunct}
{\mcitedefaultendpunct}{\mcitedefaultseppunct}\relax
\EndOfBibitem
\bibitem[Unke \latin{et~al.}(2021)Unke, Chmiela, Sauceda, Gastegger, Poltavsky,
  Schütt, Tkatchenko, and Müller]{unke_machine_2021}
Unke,~O.~T.; Chmiela,~S.; Sauceda,~H.~E.; Gastegger,~M.; Poltavsky,~I.;
  Schütt,~K.~T.; Tkatchenko,~A.; Müller,~K.-R. Machine {Learning} {Force}
  {Fields}. \emph{Chemical Reviews} \textbf{2021}, \emph{121},
  10142--10186\relax
\mciteBstWouldAddEndPuncttrue
\mciteSetBstMidEndSepPunct{\mcitedefaultmidpunct}
{\mcitedefaultendpunct}{\mcitedefaultseppunct}\relax
\EndOfBibitem
\bibitem[Jain \latin{et~al.}(2013)Jain, Ong, Hautier, Chen, Richards, Dacek,
  Cholia, Gunter, Skinner, Ceder, and Persson]{jain_commentary_2013}
Jain,~A.; Ong,~S.~P.; Hautier,~G.; Chen,~W.; Richards,~W.~D.; Dacek,~S.;
  Cholia,~S.; Gunter,~D.; Skinner,~D.; Ceder,~G.; Persson,~K.~A. Commentary:
  {The} {Materials} {Project}: {A} materials genome approach to accelerating
  materials innovation. \emph{APL Materials} \textbf{2013}, \emph{1},
  011002\relax
\mciteBstWouldAddEndPuncttrue
\mciteSetBstMidEndSepPunct{\mcitedefaultmidpunct}
{\mcitedefaultendpunct}{\mcitedefaultseppunct}\relax
\EndOfBibitem
\bibitem[Riebesell \latin{et~al.}(2023)Riebesell, Goodall, Benner, Chiang,
  Deng, Ceder, Asta, Lee, Jain, and Persson]{riebesell_matbench_2023}
Riebesell,~J.; Goodall,~R. E.~A.; Benner,~P.; Chiang,~Y.; Deng,~B.; Ceder,~G.;
  Asta,~M.; Lee,~A.~A.; Jain,~A.; Persson,~K.~A. Matbench {Discovery} -- {A}
  framework to evaluate machine learning crystal stability predictions. 2023;
  \url{https://arxiv.org/abs/2308.14920}, Version Number: 3\relax
\mciteBstWouldAddEndPuncttrue
\mciteSetBstMidEndSepPunct{\mcitedefaultmidpunct}
{\mcitedefaultendpunct}{\mcitedefaultseppunct}\relax
\EndOfBibitem
\bibitem[Yang \latin{et~al.}(2024)Yang, Hu, Zhou, Liu, Shi, Li, Li, Chen, Chen,
  Zeni, Horton, Pinsler, Fowler, Zügner, Xie, Smith, Sun, Wang, Kong, Liu,
  Hao, and Lu]{yang_mattersim_2024}
Yang,~H. \latin{et~al.}  {MatterSim}: {A} {Deep} {Learning} {Atomistic} {Model}
  {Across} {Elements}, {Temperatures} and {Pressures}. 2024;
  \url{https://arxiv.org/abs/2405.04967}, Version Number: 2\relax
\mciteBstWouldAddEndPuncttrue
\mciteSetBstMidEndSepPunct{\mcitedefaultmidpunct}
{\mcitedefaultendpunct}{\mcitedefaultseppunct}\relax
\EndOfBibitem
\bibitem[Blöchl(1994)]{blochl_projector_1994}
Blöchl,~P.~E. Projector augmented-wave method. \emph{Physical Review B}
  \textbf{1994}, \emph{50}, 17953--17979\relax
\mciteBstWouldAddEndPuncttrue
\mciteSetBstMidEndSepPunct{\mcitedefaultmidpunct}
{\mcitedefaultendpunct}{\mcitedefaultseppunct}\relax
\EndOfBibitem
\bibitem[Perdew \latin{et~al.}(1996)Perdew, Burke, and
  Ernzerhof]{perdew_generalized_1996}
Perdew,~J.~P.; Burke,~K.; Ernzerhof,~M. Generalized {Gradient} {Approximation}
  {Made} {Simple}. \emph{Physical Review Letters} \textbf{1996}, \emph{77},
  3865--3868\relax
\mciteBstWouldAddEndPuncttrue
\mciteSetBstMidEndSepPunct{\mcitedefaultmidpunct}
{\mcitedefaultendpunct}{\mcitedefaultseppunct}\relax
\EndOfBibitem
\bibitem[Kresse and Furthmüller(1996)Kresse, and
  Furthmüller]{kresse_efficiency_1996}
Kresse,~G.; Furthmüller,~J. Efficiency of ab-initio total energy calculations
  for metals and semiconductors using a plane-wave basis set.
  \emph{Computational Materials Science} \textbf{1996}, \emph{6}, 15--50\relax
\mciteBstWouldAddEndPuncttrue
\mciteSetBstMidEndSepPunct{\mcitedefaultmidpunct}
{\mcitedefaultendpunct}{\mcitedefaultseppunct}\relax
\EndOfBibitem
\bibitem[Kresse and Joubert(1999)Kresse, and Joubert]{kresse_ultrasoft_1999}
Kresse,~G.; Joubert,~D. From ultrasoft pseudopotentials to the projector
  augmented-wave method. \emph{Physical Review B} \textbf{1999}, \emph{59},
  1758--1775\relax
\mciteBstWouldAddEndPuncttrue
\mciteSetBstMidEndSepPunct{\mcitedefaultmidpunct}
{\mcitedefaultendpunct}{\mcitedefaultseppunct}\relax
\EndOfBibitem
\bibitem[Onofrio \latin{et~al.}(2015)Onofrio, Guzman, and
  Strachan]{onofrio_atomic_2015}
Onofrio,~N.; Guzman,~D.; Strachan,~A. Atomic origin of ultrafast resistance
  switching in nanoscale electrometallization cells. \emph{Nature Materials}
  \textbf{2015}, \emph{14}, 440--446\relax
\mciteBstWouldAddEndPuncttrue
\mciteSetBstMidEndSepPunct{\mcitedefaultmidpunct}
{\mcitedefaultendpunct}{\mcitedefaultseppunct}\relax
\EndOfBibitem
\bibitem[Yang \latin{et~al.}(2012)Yang, Gao, Gaba, Chang, Pan, and
  Lu]{yang_observation_2012}
Yang,~Y.; Gao,~P.; Gaba,~S.; Chang,~T.; Pan,~X.; Lu,~W. Observation of
  conducting filament growth in nanoscale resistive memories. \emph{Nature
  Communications} \textbf{2012}, \emph{3}, 732\relax
\mciteBstWouldAddEndPuncttrue
\mciteSetBstMidEndSepPunct{\mcitedefaultmidpunct}
{\mcitedefaultendpunct}{\mcitedefaultseppunct}\relax
\EndOfBibitem
\bibitem[Lee \latin{et~al.}(2023)Lee, Kim, Jeong, Kwon, Cha, Choi, Kim, and
  Jeong]{lee_electrical_2023}
Lee,~J.; Kim,~M.~J.; Jeong,~B.~G.; Kwon,~C.; Cha,~Y.; Choi,~S.~H.; Kim,~K.~K.;
  Jeong,~M.~S. Electrical role of sulfur vacancies in {MoS2}: {Transient}
  current approach. \emph{Applied Surface Science} \textbf{2023}, \emph{613},
  155900\relax
\mciteBstWouldAddEndPuncttrue
\mciteSetBstMidEndSepPunct{\mcitedefaultmidpunct}
{\mcitedefaultendpunct}{\mcitedefaultseppunct}\relax
\EndOfBibitem
\bibitem[Li \latin{et~al.}(2017)Li, Long, Bertolini, and
  Prezhdo]{li_sulfur_2017}
Li,~L.; Long,~R.; Bertolini,~T.; Prezhdo,~O.~V. Sulfur {Adatom} and {Vacancy}
  {Accelerate} {Charge} {Recombination} in {MoS}$_{\textrm{2}}$ but by
  {Different} {Mechanisms}: {Time}-{Domain} {Ab} {Initio} {Analysis}.
  \emph{Nano Letters} \textbf{2017}, \emph{17}, 7962--7967\relax
\mciteBstWouldAddEndPuncttrue
\mciteSetBstMidEndSepPunct{\mcitedefaultmidpunct}
{\mcitedefaultendpunct}{\mcitedefaultseppunct}\relax
\EndOfBibitem
\bibitem[Henkelman \latin{et~al.}(2000)Henkelman, Uberuaga, and
  Jónsson]{henkelman_climbing_2000}
Henkelman,~G.; Uberuaga,~B.~P.; Jónsson,~H. A climbing image nudged elastic
  band method for finding saddle points and minimum energy paths. \emph{The
  Journal of Chemical Physics} \textbf{2000}, \emph{113}, 9901--9904\relax
\mciteBstWouldAddEndPuncttrue
\mciteSetBstMidEndSepPunct{\mcitedefaultmidpunct}
{\mcitedefaultendpunct}{\mcitedefaultseppunct}\relax
\EndOfBibitem
\bibitem[Komsa and Krasheninnikov(2015)Komsa, and
  Krasheninnikov]{komsa_native_2015}
Komsa,~H.-P.; Krasheninnikov,~A.~V. Native defects in bulk and monolayer {MoS}
  2 from first principles. \emph{Physical Review B} \textbf{2015}, \emph{91},
  125304\relax
\mciteBstWouldAddEndPuncttrue
\mciteSetBstMidEndSepPunct{\mcitedefaultmidpunct}
{\mcitedefaultendpunct}{\mcitedefaultseppunct}\relax
\EndOfBibitem
\bibitem[Noh \latin{et~al.}(2014)Noh, Kim, and Kim]{noh_stability_2014}
Noh,~J.-Y.; Kim,~H.; Kim,~Y.-S. Stability and electronic structures of native
  defects in single-layer {MoS} 2. \emph{Physical Review B} \textbf{2014},
  \emph{89}, 205417\relax
\mciteBstWouldAddEndPuncttrue
\mciteSetBstMidEndSepPunct{\mcitedefaultmidpunct}
{\mcitedefaultendpunct}{\mcitedefaultseppunct}\relax
\EndOfBibitem
\bibitem[Freysoldt \latin{et~al.}(2014)Freysoldt, Grabowski, Hickel,
  Neugebauer, Kresse, Janotti, and Van
  De~Walle]{freysoldt_first-principles_2014}
Freysoldt,~C.; Grabowski,~B.; Hickel,~T.; Neugebauer,~J.; Kresse,~G.;
  Janotti,~A.; Van De~Walle,~C.~G. First-principles calculations for point
  defects in solids. \emph{Reviews of Modern Physics} \textbf{2014}, \emph{86},
  253--305\relax
\mciteBstWouldAddEndPuncttrue
\mciteSetBstMidEndSepPunct{\mcitedefaultmidpunct}
{\mcitedefaultendpunct}{\mcitedefaultseppunct}\relax
\EndOfBibitem
\bibitem[Prasad \latin{et~al.}(2023)Prasad, Al-Ani, Goss, and
  Mar]{prasad_charge_2023}
Prasad,~M.~K.; Al-Ani,~O.~A.; Goss,~J.~P.; Mar,~J.~D. Charge transfer due to
  defects in hexagonal boron nitride/graphene heterostructures: {An} \textit{ab
  initio} study. \emph{Physical Review Materials} \textbf{2023}, \emph{7},
  094003\relax
\mciteBstWouldAddEndPuncttrue
\mciteSetBstMidEndSepPunct{\mcitedefaultmidpunct}
{\mcitedefaultendpunct}{\mcitedefaultseppunct}\relax
\EndOfBibitem
\end{mcitethebibliography}

\end{document}